\documentclass[a4paper,11pt,oneside ]{article}

\usepackage{jheppub}
\usepackage[T1]{fontenc} 
\usepackage[utf8]{inputenc}
\usepackage{physics}
\usepackage{xcolor}
\usepackage{mathrsfs}
\usepackage{tikz}
\usepackage{colortbl}
\def\mpl{M_{\rm Pl}}
\allowdisplaybreaks
\newcommand{\nc}{\newcommand}
\nc{\fNL}{{f_{\text{NL}}}}
\nc{\dfNL}{{f'_{\text{NL}}}}

\usepackage[framemethod=TikZ]{mdframed}
\mdfsetup{roundcorner=5pt}

\newmdenv[
skipabove=5pt,
skipbelow=7pt,
rightline=false,
leftline=false,
topline=false,
bottomline=false,
backgroundcolor=gray!30,
innerleftmargin=5pt,
innerrightmargin=5pt,
innertopmargin=5pt,
innerbottommargin=5pt,
leftmargin=0cm,
rightmargin=0cm,
linewidth=4pt]{eBox}

\hypersetup{colorlinks=true,urlcolor=pacificblue,citecolor=magenta,linkcolor=blue}

\graphicspath{{figures/}}

\definecolor{pacificblue}{cmyk}{0.95,0.3,0, 0.5}
\definecolor{Blue}{HTML}{2171b5}

\nc{\ba}{\begin{eqnarray}}
\nc{\ea}{\end{eqnarray}}
\newcommand{\calR}{{\cal{R}}}

\newcommand{\calP}{{\cal{P}}}

\nc{\reh}{{\text{reh}}}
\nc{\bfx}{{\bf{x}}}
\nc{\bfk}{{\bf{k}}}
\nc{\mhn}{\color{magenta}{\bf MHN: }}
\nc{\magenta}{\color{magenta}{ }}

\makeatletter
\gdef\@fpheader{}
\makeatother

\title{Non-Gaussianity consistency relations \\ and their consequences for the peaks}

\abstract{
Strong deviations from scale invariance and the appearance of high peaks in the primordial power spectrum have been extensively studied for generating primordial black holes (PBHs) or gravitational waves (GWs). It is also well-known that the effect of non-linearities can be significant in both phenomena. In this paper, we advocate the existence of a general single-field consistency relation that relates the amplitude of non-Gaussianity in the squeezed limit $f_{\text{NL}}$ to the power spectrum and remains valid when almost all other consistency relations are violated. In particular, it is suitable for studying scenarios where scale invariance is strongly violated.
We discuss the general and model-independent consequences of the consistency relation on the behavior of $f_{\text{NL}}$ at different scales.
Specifically, we study the size, sign and slope of $f_{\text{NL}}$ at the scales where the power spectrum peaks and argue that generally the peaks of $f_{\text{NL}}$ and the power spectrum occur at different scales. 
As an implication of our results, we argue that non-linearities can shift or extend the range of scales responsible for the production of PBHs or GWs, relative to the window as determined by the largest peak of the power spectrum, and may also open up new windows for both phenomena.
}

\begin{document}
	
	\author{Mohammad Hossein Namjoo,}
	\author{Bahar Nikbakht}
	
	\affiliation{School of Astronomy, Institute for Research in Fundamental Sciences (IPM), \\ Tehran, Iran, P.O. Box 19395-5531}
	
	\emailAdd{mh.namjoo@ipm.ir}
	\emailAdd{bahar.nikbakht@ipm.ir}
	
\maketitle
\flushbottom

\section{Introduction}
\label{sec:intro}
The possibility of large enhancement of power at sub-CMB scales has been widely studied due to its necessity for the abundant formation of the primordial black holes (PBHs)  \cite{Ozsoy:2023ryl,Hertzberg:2017dkh,Garcia-Bellido:2017mdw} and the generation of induced gravitational waves (GWs) \cite{Domenech:2021ztg}. Such large enhancements in the power spectrum can occur if  scale invariance is broken during inflation as a result of phase transitions and temporary phases of non-attractor evolution. It is also well-known that  the PBH formation and the generation of GWs are both sensitive to the non-linearities that may be represented by higher order correlation functions \cite{Bullock:1996at,Cai:2018dig,Young:2015kda,Atal:2018neu,Franciolini:2023pbf}. Thus, it is crucial to simultaneously study the power spectrum and non-linearities at the scales where the enhancements occur. This paper provides one step toward that direction by paying particular attention to the scale-dependence of the two-point and three-point correlation functions. More explicitly, we study the squeezed limit bispectrum which corresponds to the correlation between the modes that experience large enhancements and a mode that has a much larger wavelength.

A useful tool that may enable us to simultaneously study the power spectrum and the squeezed limit bispectrum is the non-Gaussianity consistency relations.  However, the aforementioned reasons for the enhancement of the power spectrum --- i.e., the presence of phase transitions, the break-down of scale invariance and the non-attractor phases of inflation --- are in fact different reasons for the violation of almost all non-Gaussianity consistency relations, if the long mode is affected by these features \cite{Finelli:2017fml,Hinterbichler:2012nm,Bravo:2017wyw,Assassi:2012zq,Cheung:2007sv,Kundu:2015xta}. An exception is the consistency condition that is advocated in Ref.~\cite{Namjoo_2023} which is claimed to hold more generally because it does not rely on specific symmetries or restrictive assumptions about the Lagrangian governing inflation. In this paper, we confirm that claim by customizing the consistency relation of Ref.~\cite{Namjoo_2023} specifically for the situation of interest, which is the single-field inflationary scenarios with possibly multiple phases of non-attractor evolution and sharp phase transitions that lead to the large enhancement of the power spectrum and the appearance of several peaks and troughs (Sec.~\ref{sec:general_cosistency}). 

If the long mode is not affected by the aforementioned features in the inflationary background --- which happens if the scale-invariant-breaking phases are sufficiently short --- the slow-roll consistency relation \cite{Maldacena_2003} holds. This relation implies that the size of the squeezed limit bispectrum is determined by the degree to which the power spectrum deviates from scale invariance. Thus, there can be a strong correlation between the long mode and the short mode due to the strong violation of scale invariance. We will show that our general consistency relation reduces to the standard one in the same limit (Sec.~\ref{sec:slow-roll_consistency}). 

Since our consistency relation is generic, we can make general and model-independent statements about the behavior of the squeezed limit bispectrum as a function of the wavelength of the short modes. In particular, we study the size, sign and slope of the bispectrum at the scales where the power spectrum peaks (Sec.~\ref{sec:peaks}). Moreover, we show that the peaks of the power spectrum and the squeezed limit bispectrum generally occur at different scales. Crucially, this implies that the largeness of the power spectrum and the significance of non-linearities may contribute to the PBH formation separately at different scales, changing the mass window of PBHs relative to the predictions of the linear theory. Likewise, we argue that the scales where non-linearities can have a notable impact on the generation of GWs  may differ from the peak scale of the power spectrum. We confirm the validity of the consistency relation and its implications using a concrete example (Sec.~\ref{sec:example}).

In the rest of the introduction, we provide some definitions and clarify our notation. We work with the curvature perturbation on comoving hypersurfaces, denoted by $\mathcal{R}$. Generally, its two-point, unequal-time correlation function may be quantified by
\begin{equation}
\label{eq:power spectrum}
\left\langle\mathcal{R}_{\mathbf{k}}(\tau) \mathcal{R}_{\mathbf{k'}}(\tau')\right\rangle = (2 \pi )^3 \delta^{(3)}\left(\mathbf{k}+\mathbf{k'}\right) P_{\mathcal{R}_{k}}(\tau,\tau').
 \end{equation}
The observable correlation function is the equal-time at the boundary (i.e., at the end of inflation, $\tau\to 0$) which we denote by $P_{\mathcal{R}_{k}}$ (without an explicit time argument). 
Note that,  $P_{\mathcal{R}_{k}}(\tau,\tau')$ is not necessarily positive or even real, unless if $\tau'=\tau$. However, we will need this notation only for the long-wavelength, super-horizon mode which is classical, resulting in the reality of that function. It will be useful to express the results in terms of the dimensionless power spectrum  $\mathcal{P}_{\mathcal{R}_k}(\tau,\tau') \equiv \frac{k^3}{2 \pi^2} P_{\mathcal{R}_k}(\tau,\tau')$ and $\mathcal{P}_{\mathcal{R}_k}\equiv \frac{k^3}{2 \pi^2} P_{\mathcal{R}_k}$ if $\tau=\tau'=0$.  

The three-point function at the boundary, the bispectrum, may be quantified by
\begin{equation}
    \label{eq:bispectrum}
    \left\langle\mathcal{R}_{\mathbf{k}_{\mathbf{1}}} \mathcal{R}_{\mathbf{k}_{\mathbf{2}}} \mathcal{R}_{\mathbf{k}_{\mathbf{3}}}\right\rangle \equiv(2 \pi)^3 \delta^{(3)}\left(\mathbf{k}_{\mathbf{1}}+\mathbf{k}_{\mathbf{2}}+\mathbf{k}_{\mathbf{3}}\right) B_{\mathcal{R}}\left(k_1, k_2, k_3\right) .
\end{equation}
We will be interested in the squeezed limit bispectrum where the momentum of one mode (which we denote by $k_\ell$) is much smaller than the momentum of the other two (which we denote by $k_s$). We quantify the bispectrum in this limit by 
\begin{equation}
    \label{eq:fnl}
    B_{\mathcal{R}}\left(k_\ell, k_s, k_s\right) \to (2 \pi)^4 \frac{1}{k_s^3 k_\ell^3} \mathcal{P}_{\mathcal{R}_\ell} \mathcal{P}_{\mathcal{R}_s} \, \frac{3}{5} \fNL(k_s)\, , \qquad \text{for $k_\ell \ll k_s$}\ ,
\end{equation}
where we used $\mathcal{R}_\ell$ and $\mathcal{R}_s$ as shorthands for $\mathcal{R}_{k_\ell}$ and $\mathcal{R}_{k_s}$, respectively. Eq.~\eqref{eq:fnl} also defines  $\fNL$ as the amplitude of non-Gaussianity which in general depends on the wavenumber $k_s$. When a relation applies to both long and short modes, we denote the corresponding wavenumber by $k$. Occasionally, we will encounter the ratio of the power spectra of the long mode at different times. Thus, we find it useful to also define 
\ba 
\label{eq:r}
r(\tau) \equiv \calP_{\calR_\ell}(0,\tau)/\calP_{\calR_\ell}.
\ea 

To simplify the language, when confusion is not possible, we refer to the scales where the power spectrum peaks as the {\it  peak scale} or the {\it  peak}. Likewise, we refer to the scale where there is a trough in the power spectrum as the {\it  trough scale} or the {\it trough}. We may also be interested in the peaks of other quantities but we refer to them explicitly to avoid ambiguities. 
Finally, throughout this paper, we set $\mpl\equiv (8\pi G)^{-1/2}=1$.

\section{A general consistency relation} 
\label{sec:general_cosistency}
A  general consistency relation for single-field inflation is presented in Ref.~\cite{Namjoo_2023} where it is shown that it reproduces the results of many single-field inflationary models including the ones that contain a non-attractor phase of inflation. Here, we review the consistency condition of Ref.~\cite{Namjoo_2023} and customize it to be particularly convenient for the analysis of the models of interest where some modes experience phases of non-attractor evolution leading to the strong violation of  scale invariance.

The first step in the derivation of the consistency relation is to note that the correlation between the long mode (which is assumed to be a super-horizon, classical fluctuation) and the short modes can be written by \cite{Creminelli:2004yq}
\ba 
\label{long_short_3pt}
\langle \calR_\ell(\tau) \calR_s(\tau)  \calR_s(\tau) \rangle \simeq \langle \calR_\ell(\tau)\, \langle \calR_s(\tau)  \calR_s(\tau)\rangle_{\calR_\ell} \rangle \ ,
\ea 
where $\langle \calR_s^2 \rangle_{\calR_\ell}$ denotes the two-point correlation function of the short modes under the influence of the long mode and $\tau$ is the conformal time at which the correlations are computed which, from now on, we set $\tau \to 0$. According to Eq.~\eqref{long_short_3pt}, the long-short correlation arises from the short modes perceiving the long mode as a change in their background evolution. This intuition is consistent with the separate universe picture but note that we need not to assume that $\calR_s$ is a super-horizon mode when it starts correlating with the long mode. Assuming that the effect of the long mode is small, Eq.~\eqref{long_short_3pt} can be Taylor expanded. Since we do not necessarily require $\calR_\ell$ to be a constant in time, its effect must be taken into account over the entire history of the short modes. Thus, we perform a {\it functional Taylor expansion} as follows
\ba 
\label{eq:Taylor}
\langle \calR_s^2(0) \rangle_{\calR_\ell} \simeq \langle \calR_s^2(0) \rangle_{0} + \int_{\tau_\ell}^{0} \calR_\ell(\tilde \tau) \dfrac{\delta }{\delta \calR_\ell(\tilde \tau) } \langle \calR_s^2(0) \rangle \, d\tilde  \tau \, ,
\ea 
where  higher order terms in the expansion are neglected. $\tau_\ell$ is around the horizon-crossing time of $\calR_\ell$ and we assumed that all correlations between the long mode and the short modes occur after $\tau_\ell$. This is justified as long as the sub-horizon short modes around $\tau_\ell$ are in the Bunch-Davies vacuum. Note that, the right hand side of Eq.~\eqref{eq:Taylor} being a Taylor expansion, we must set $\calR_\ell=0$ in $\delta \langle \calR_s^2 \rangle / \delta \calR_\ell$ after the functional derivative is taken.  Substituting this relation to Eq.~\eqref{long_short_3pt} results in 
\ba 
\label{main}
B(k_s,k_s,k_\ell) \simeq  \int_{\tau_\ell}^{0} \langle \calR_\ell(0) \calR_\ell(\tilde \tau) \rangle' \dfrac{\delta }{\delta \calR_\ell(\tilde \tau) } \langle \calR_s^2(0) \rangle'  \, d\tilde \tau \, ,
\ea 
where $\langle . \rangle'$ denotes $\langle . \rangle$ with the factor $(2\pi)^3$ and the momentum conserving delta function dropped.
The notation introduced in Sec.~\ref{sec:intro} (in particular, the definition $r(\tau)$ according to Eq.~\eqref{eq:r}) allows us to also write 
\ba 
\label{main_fNL}
\dfrac{12}{5} \fNL =  \int_{\tau_\ell}^{0} r(\tilde \tau)\, \dfrac{\delta \ln \calP_{\calR_s}}{\delta \calR_\ell(\tilde \tau) } \, d\tilde \tau. 
\ea 
This is the key relation that is shown in Ref.~\cite{Namjoo_2023} that can reproduce the squeezed limit bispectra from many single-field models of inflation. We refer the readers to Ref.~\cite{Namjoo_2023} for a summary of the assumptions behind this relation and for its application to some non-trivial examples where many other consistency relations are violated. Here we only mention that, besides the conditions stated above, this relation assumes that  the long mode can be described by a single independent solution. That is why the functional derivatives are taken with respect to $\calR_\ell$ and not its conjugate momentum. In most situations, this assumption is satisfied (even if $\calR_\ell$ is time dependent) but the extension of Eq.~\eqref{main} to also include the effect of the conjugate momentum is straightforward. To utilize Eq.~\eqref{main_fNL}, one needs to inspect how the late time power spectrum of the short mode  $\calP_{\calR_s}$ depends on the history of the evolution and then study how this history may be affected by a long wavelength mode. In practice, $\calP_{\calR_s}$ depends on a few instances of time in its history leading to a significant simplification. 
 
 \subsection{General consistency relation for non-scale-invariant correlations}
 One can stop with the abstract formula Eq.~\eqref{main_fNL} and apply it to specific examples. However, to reach some generic conclusions for the scenarios that are of  interest in this paper, we make further --- but still generic --- assumptions  to simplify Eq.~\eqref{main_fNL}. What we have in mind is inflationary scenarios with transient phases of non-attractor evolution that result in a large enhancement of the power spectrum at sub-CMB scales. The non-attractor phases can happen e.g., due to the presence of  constant segments in the potential (the so-called ultra-slow-roll phase) or breaks in the potential or its derivative. The transitions from a short non-attractor phase to the subsequent slow-roll phase must be sharp enough so that the effect of the non-attractor evolution on the power spectrum is not washed out \cite{Cai_2018}. Then one expects to have a large peak followed by several (usually smaller) ones in the power spectrum.\footnote{Obviously, when there are peaks, troughs would also exist but peaks are of more interest due to their dominant contribution to the PBH or GW production.} Our goal is to understand how the squeezed limit bispectrum behaves at the scales where the power spectrum deviates significantly from scale invariance and, specifically, at the peaks. 
 
Assume that $\calP_{\calR_s}$ is an explicit function of several instances of time and the background initial conditions, i.e.,  $\calP_{\calR_s} = \calP_{\calR_s}(k_s,\phi_{0},\tau_*, \tau_0, \tau_1,...,\tau_{n})$, where $\tau_0$ is some initial conformal time\footnote{$\tau_0$ must be early enough so that the effect of  even earlier history on $\calP_{\calR_s}$ can be neglected but it also needs to be late enough so that the long mode has already left the horizon (i.e., $|\tau_0| \lesssim |\tau_\ell|$).}, $\phi_0$ depicts collectively the dependence on the initial conditions (i.e., the value of the inflaton field and its conjugate momentum at $\tau_0$) and we have also taken into account a possible dependence on the horizon-crossing time $\tau_*=-1/k_s$.\footnote{We assume that all other conformal times in the argument of $\calP_{\calR_s}$ are independent of $k_s$.} We do not make explicit assumptions about the other instants  $\tau_j$ for $j=1,...,n$ but they can appear e.g., when $n$ distinct phase transitions occur, responsible for the large enhancement of the short mode power spectrum.  To derive a more explicit expression for $\fNL$ from Eq.~\eqref{main_fNL}, we need to inspect how the presence of the long mode affects $\calP_{\calR_s}$ due to its effect on the history of $\calR_s$. First, note that the long mode induces an initial perturbation to the inflaton field (and thus to its conjugate momentum) via $\Delta \phi_0 = -(\dot \phi_0/H) \calR_\ell(\tau_0)$. Furthermore, the line element as seen by the short mode in the presence of the long mode is given by $ds^2\simeq -dt^2 + a^2 e^{2\calR_\ell} d\bfx^2$. Therefore, the effect of $\calR_\ell$ on different instances of time can be described by a rescaling of the scale factor, $a\to \hat  a = a e^{\calR_\ell}$.  A positive $\calR_\ell$ leads to a larger scale factor, which implies that the conformal time of each instant which $\calP_{\calR_s}$ depends on moves forward. One exception is the horizon-crossing conformal time which must remain fixed for a fixed comoving wavenumber $k_s$ which implies  that a positive $\calR_\ell$ shifts $\tau_*$ backward. Taking all these effects into account results in 
 \ba
 \dfrac{\delta \ln \calP_{\calR_s}}{\delta \calR_\ell(\tilde \tau) } =   \dfrac{\partial \phi_0}{\partial \calR_\ell} \dfrac{\partial \ln \calP_{\calR_s}}{\partial \phi_0} \delta(\tilde \tau-\tau_0) +    \dfrac{d \ln \calP_{\calR_s}}{d \ln \tau_*} \delta(\tilde \tau-\tau_*) - \sum_{j=0}^{n}   \dfrac{d \ln \calP_{\calR_s}}{d \ln \tau_j} \delta(\tilde \tau-\tau_j)\ ,
 \ea 
 where the first term is schematic and takes into account all effects of $\calR_\ell$ on the initial conditions. 
 Plugging this into Eq.~\eqref{main_fNL} results in
 \ba 
 \label{eq:fNL_interm}
 \dfrac{12}{5} \fNL  = r(\tau_0)\, \dfrac{\partial \phi_0}{\partial \calR_\ell} \dfrac{\partial \ln \calP_{\calR_s}}{\partial \phi_0}  + r(\tau_*)  \dfrac{d \ln \calP_{\calR_s}}{d \ln \tau_*}  
  - \sum_{j=0}^{n}  r(\tau_j) \dfrac{d \ln \calP_{\calR_s}}{d \ln \tau_j}.
 \ea 
 To attain some insight from this key result,  we note that the characteristic comoving scales are only meaningful relative to each other.  Thus, defining $x_{k_s}=-k_s\tau_n$, and $x_j = \tau_j/\tau_n$ for $0 \leq j <n$,  the power spectrum can be described by $\calP_{\calR_s} = \calP_{\calR_s}(\phi_0, \tau_*,x_{k_s}, x_1,...,x_{n-1})$ where we kept the dependence on $\tau_*$ explicit due to its different role and meaning. This allows us to rewrite Eq.~\eqref{eq:fNL_interm} as 
 \ba 
 \label{eq:fNL_final}
 \nonumber 
 \dfrac{12}{5} \fNL  &=&  
r(\tau_0)\, \dfrac{\partial \phi_0}{\partial \calR_\ell} \dfrac{\partial \ln \calP_{\calR_s}}{\partial \phi_0}  
 +  \sum_{j=0}^{n-1} \big( r(\tau_n) - r(\tau_j)\big) \dfrac{\partial \ln \calP_{\calR_s}}{\partial \ln x_j}
 \\
 &+&\big( r(\tau_*) - r(\tau_n)\big) \dfrac{d\ln \calP_{\calR_s}}{d \ln \tau_*}  
 -r(\tau_n) (n_s-1) \, ,
 \ea 
 where  $(n_s-1)$ is the spectral index that takes into account both the explicit and implicit dependence on $k_s$:
 \ba 
n_s(k_s)-1\equiv \dfrac{d \ln \calP_{\calR_s}}{d\ln k_s} = x_{k_s} \dfrac{\partial \ln \calP_{\calR_s}}{\partial x_{k_s}} - \tau_* \dfrac{\partial \ln \calP_{\calR_s}}{\partial  \tau_*}.
 \ea 
Eq.~\eqref{eq:fNL_interm} (or Eq.~\eqref{eq:fNL_final}) is the final form of $\fNL$ that, we claim, reproduces the bispectrum of all examples for which our assumptions are satisfied (including many models studied in the literature for the purpose of generating PBHs or GWs). This expression may be further  simplified when further assumptions are made or a more explicit form of $\calP_{\calR_s}$ is considered. In the rest of this section and in Sec.~\ref{sec:peaks}, we discuss the consequences of this generic result  and in Sec.~\ref{sec:example} we study a concrete example to examine our consistency relation.
 
 \subsection{Conserved long mode and the transient non-attractor evolution}
 \label{sec:slow-roll_consistency}
First, we consider a situation where the long mode $\calR_\ell$ has left the horizon during the slow-roll inflation prior to any non-attractor phase and is completely frozen when the short modes experience non-attractor evolution. The attractor behavior is expected to be reached quickly due to the rapid Hubble expansion, making the non-attractor phases of inflation very short. Therefore, this situation indeed covers most realistic scenarios that generate large peaks in the power spectrum. In this case, the initial time $\tau_0$  is in the slow-roll phase well before the onset of the non-attractor evolution. This implies that the power spectrum is insensitive to the initial conditions and the first term of Eq.~\eqref{eq:fNL_final} can be neglected. Moreover, $\calR_\ell$ is conserved during the entire evolution that lays down the final form of $\calP_{\calR_s}$. That is, we have $r(\tau) \equiv \calP_{\calR_\ell}(0,\tau)/\calP_{\calR_\ell}=1$ for all $\tau > \tau_\ell$ which significantly simplifies Eq.~\eqref{eq:fNL_final} and reduces it to 
  \ba
  \label{eq:fNL_SR}
 \dfrac{12}{5} \fNL = -(n_s-1)\ .
 \ea 
 Remarkably, this is precisely the standard, slow-roll consistency relation (but holds even if the short modes --- but not the long mode --- experience non-attractor evolution) \cite{Maldacena_2003}. This result is  in fact expected since this consistency relation only relies on the conservation of  the long mode and does not require restrictions on the evolution of the short modes. 
 
 It is well-known that the consistency relation Eq.~\eqref{eq:fNL_SR} can be violated if the long mode is not conserved which can happen if the long mode also experiences non-attractor phases of inflation \cite{Namjoo:2012aa}.   
 Our general consistency relation Eq.~\eqref{eq:fNL_final} holds even in this case since the conservation of the long mode is not required. 
 
 \subsection{ Bispectrum at  large and short scales}
 \label{sec:limits}
 If the mode of interest $\calR_s$ experiences non-attractor phases when its wavelength is larger than the horizon, the resulting changes to the correlation functions will be scale-invariant. This is the situation where the methods that rely on the gradient expansion for all modes, such as the $\delta N$-formalism, are expected to hold. Because of the scale invariance, the third and the last terms  in Eq.~\eqref{eq:fNL_final} vanish and we are left with the dependence on the initial conditions and the relative time scales of the non-attractor phases. We will see in Sec.~\ref{sec:example} that our consistency relation reproduces the results from the $\delta N$-formalism at these scales (while our result is more generic and works for all scales). 
 
Eq.~\eqref{eq:fNL_final}  allows us to also make  generic statements about $\fNL$ at very short scales $k_s \to \infty$. As discussed in Sec.~\ref{sec:general_cosistency}, the scenarios of interest are expected to have sharp transitions. Consequently, at short scales the power spectrum oscillates rapidly. Because of the strong violation of scale invariance and the rapid oscillations of the power spectrum, in Eq.~\eqref{eq:fNL_final} the term  proportional to the spectral index dominates and we simply have $\fNL \propto (n_s-1)$ in this limit.\footnote{Note that in realistic scenarios, the transitions are not infinitely sharp. Consequently, this argument fails if one goes to extremely short scales where the corresponding modes perceive the transition as smooth.} The proportionality constant is model-dependent and generally differs from the one in the slow-roll consistency relation Eq.~\eqref{eq:fNL_SR}. This result will have important consequences for the peak scales which will be discussed in Sec.~\ref{sec:peaks}. In the explicit example of Sec.~\ref{sec:example_extended}  we will see that $\fNL$  converges  quickly to this result as one goes to short scales and strong deviations from it only occur at the first few peaks.

\section{Consequences of the consistency relation for the peaks}
\label{sec:peaks}
Having equipped with the general consistency relation, in this section we make general statements about the size, sign and slope of the squeezed limit bispectrum at the peak scales.  What we obtain here is of significant importance concerning PBHs or GWs production.

\subsection{Size of bispectrum}
\label{sec:size}
Concerning the size of $\fNL$, the most realistic situation --- i.e., when the non-attractor phase is short --- is indeed the simplest one to analyze because of the simplicity of the consistency relation Eq.~\eqref{eq:fNL_SR}. The simple conclusion is that $\fNL$ vanishes at the peaks, where $d\calP_{\calR_s}/dk \propto (n_s-1)=0$.

When Eq.~\eqref{eq:fNL_SR} is violated,  the more general form of consistency relation Eq.~\eqref{eq:fNL_final} holds which allows for large values of $\fNL$ at the peaks. We will confirm this in an explicit example in Sec.~\ref{sec:example_extended}. However, we note that when we go to scales shorter than the first peak --- as discussed in Sec.~\ref{sec:limits} --- $\fNL$ will be dominated by a  term proportional to $(n_s-1)$. This implies that scales away from the peaks may have much larger values of $\fNL$. This may cause the realization of large fluctuations to become more probable  at  non-peak scales which will be explored in the explicit example of Sec.~\ref{sec:example} in Secs.~\ref{sec:beta} and \ref{sec:GW}.

\subsection{Sign of bispectrum}
\label{sec:sign}

The sign of the squeezed limit bispectrum  is also of interest  since the rate of PBH formation  is sensitive to it \cite{Young:2015kda}. When the slow-roll consistency relation Eq.~\eqref{eq:fNL_SR} holds, $\fNL$ vanishes at the peaks so this question is not relevant. Therefore, we focus on the cases where Eq.~\eqref{eq:fNL_SR} is violated.  

The sign of $\fNL$ is studied in Ref.~\cite{firouzjahi2023sign} which claims that $\fNL$ at the peaks is always positive. We will see in the explicit example of Sec.~\ref{sec:example_extended} that $\fNL$ at the peak is different from what is obtained in Ref.~\cite{firouzjahi2023sign} and argue that the result of Ref.~\cite{firouzjahi2023sign} can indeed become negative. However, our consistency relation Eq.~\eqref{eq:fNL_final} is sufficiently powerful to allow us to extract the assumptions required for the positivity of $\fNL$ at the peaks (besides the one required for the validity of Eq.~\eqref{eq:fNL_final} itself).  Recall that $\tau_n$ was an arbitrary reference time that $\calP_{\calR_s}$ depends on.  To simplify the analysis, without loss of generality, we assume that $\tau_n$ is chosen so that $r(\tau_n) \geq  r(\tau_j)$ for all $j\neq n$. 
Then it is easy to see that, according to Eq.~\eqref{eq:fNL_final}, $\fNL$ is positive at the peaks if (i) the dependence of $\calP_{\calR_s}$ on the horizon crossing time $\tau_*$ can be neglected, (ii) the dependence of $\calP_{\calR_s}$ on the initial conditions can be neglected, (iii) $r(\tau_n)$ (hence $\calP_{\calR_\ell}(0,\tau_n)$) is positive  and  (iv) a longer separation between $\tau_n$ and $\tau_j$ for all $j\neq n$ leads to a larger power spectrum at the peaks.\footnote{Note that a simultaneous negation of conditions (iii) and (iv) also leads to the same conclusion, i.e., the positivity of $\fNL$ at the peak. However, this does not happen --- at least --- for the simplest non-attractor models of inflation.} We do not claim that when any of the above assumptions is violated $\fNL$ will or even can be negative. If there is a more general theorem stating the positivity of $\fNL$ beyond these assumptions, it remains to be proven. Assumptions (i) and (ii) are satisfied in many scenarios that predict large peaks in the power spectrum.\footnote{See, however, the discussion in Ref.~\cite{Namjoo_2023} where it is shown that the example of Ref.~\cite{Chen:2013aj} violates condition (ii).} Condition (iii) holds  if the mode function of $\calR_\ell$ does not change sign during which it affects the short modes (which is often the case for non-attractor evolutions with sharp transitions).  (iv) is also likely to be satisfied in most scenarios since $\tau_n$  is likely to be the conformal time at the final phase transition after which the attractor behavior is reached and then  condition (iv) corresponds to requiring a larger $\calP_{\calR_s}$ at the peak for a longer non-attractor period.

We conclude that despite the restrictive appearance of the above assumptions, they are indeed satisfied in many scale-invariance-breaking models of inflation that lead to large enhancement of the power spectrum, hence $\fNL$ is likely to be positive at the peaks. The complex example that will be studied in Sec.~\ref{sec:example} confirms that conclusion in the entire parameter space. It would be interesting to explore the possibility of violations of these assumptions and whether that can lead to a negative $\fNL$ at the peaks. In particular, the sensitivity to the initial conditions can happen in models like the one in Ref.~\cite{Chen:2013aj}. Also, conditions (iii) and (iv) may be violated if  a smooth phase transition occurs or if multiple phases of non-attractor evolution with sufficiently complex dynamics exist.

\subsection{Slope of bispectrum}
\label{sec:slope}
We can also study the slope of $\fNL$ around the peak scales using the consistency relations. In the scenarios where the slow-roll consistency relation Eq.~\eqref{eq:fNL_SR} holds, it is obvious that we have $\dfNL>0$ at the peaks. When Eq.~\eqref{eq:fNL_SR} is violated, from Eq.~\eqref{eq:fNL_final}  the same conclusion can be deduced if conditions (i) and (iii) of Sec.~\ref{sec:sign} are satisfied. Since these assumptions typically hold in models of interest, we conclude that the positivity of $\dfNL$ at the peaks is generic. This result may have notable implications for the PBHs and GWs. In particular, together with the positivity of $\fNL$, it shows that $\fNL$ is larger at shorter-than-the-peak scales. Thus, if  non-linearities are strong enough to shift the window of scales responsible for the generation of PBHs or GWs, the shift ought to be toward the shorter-than-the-peak scales. (See Secs.~\ref{sec:beta} and \ref{sec:GW} for related illustrative examples.)
\\

We end this section with a few  remarks on the troughs. A similar analysis can be done to investigate the size, sign and slope of the bispectrum at the trough scales. In particular, one can argue for the non-positivity of $\fNL$ and negativity of $\dfNL$ at the troughs, under the similar conditions discussed above. Crucially, in different examples that we have studied, when condition (iv) holds and the power spectrum at the peak scales increases for a longer period of non-attractor evolution, the power spectrum at the trough scales decreases. This is the reason for the negativity of $\fNL$.  However, it is worth noting that --- depending on the details of the model --- the power spectrum  at the trough scale may vanish which is the minimum value that it can take. In this case, $\fNL$ also vanishes (rather than being negative) according to the consistency condition Eq.~\eqref{eq:fNL_final}. Thus, in this sense, we are in a similar situation as where the slow-roll consistency relation Eq.~\eqref{eq:fNL_SR} holds, even though it may be strongly violated.

\section{A concrete example: Ultra-slow-roll inflation followed by a step}
\label{sec:example}
In this section, we verify the validity of the consistency relation Eq.~\eqref{eq:fNL_final} in a concrete example and make explicit its consequences for the peaks. 
We consider a transient ultra-slow-roll (USR) phase sandwiched between two slow-roll (SR) phases and also consider an infinitely sharp step in the potential at the end of the USR phase. Such a sharp step is unrealistic  but we take it into account to examine  our consistency relation in a more complex but yet tractable example. The USR phase occurs when there is a flat segment of the potential. We will have two transitions (from SR to USR and vice versa) that we assume are both instantaneous and model the potential as follows
\begin{equation}
\label{eq:Potential}
	V(\phi) =  V_0 \left[ 1 +  \left(\kappa +  \sqrt{2 \epsilon_V} (\phi-\phi_e) \right) \, \theta(\phi_e - \phi) + \sqrt{2 \epsilon_i} (\phi-\phi_i)\, \theta(\phi - \phi_i)\right],
\end{equation}
where $V_0$ is a constant (and the dominant term), $\epsilon_i$ and $\epsilon_V$ control the constant slopes of the slow-roll pieces (and coincide with the attractor slow-roll parameter at each piece) and $\phi_i$ and $\phi_e$ are the field values at the SR-USR and USR-SR transitions, respectively. $\kappa$ determines the size of the step after the USR phase; if $\kappa>0$ ($\kappa<0$) we have an upward (a downward) step.  We assume $|\kappa| \ll 1$. We denote the conformal time at the two transitions by $\tau_i$ and $\tau_e$, respectively. 
 We also denote the slow-roll parameter $\epsilon \equiv -\dot H/H^2$ at the end of the USR phase by $\epsilon_e$ and right after the step by $\epsilon_s$. To better express our final results we follow Refs.~\cite{Cai:2022erk,Cai_2018} and also define 
 \ba 
 h\equiv -6\sqrt{\frac{\epsilon_V}{\epsilon_s}}\ , \qquad g \equiv \sqrt{\frac{\epsilon_s}{\epsilon_e}}.
 \ea  
 $h$ determines the relaxation time to the final slow-roll phase and $g$  quantifies the abrupt change of the slow-roll parameter due to the step. By definition, we always have $h<0$; larger values of $|h|$ correspond to the faster relaxation (after which the super-horizon $\calR$ freezes). Furthermore, we always have $g>0$; $0<g<1$ for an upward  and $g>1$ for a downward step. The step disappears when $g=1$. Since we assume that the step is instantly taken by inflaton, from the energy conservation, we have $\epsilon_s=\epsilon_e -3\kappa$.  During the USR phase the slow-roll parameter decays like $\epsilon \sim a^{-6}$ leading to the rapid growth of the  curvature perturbation like $a^{3}$. 
 
 One advantage of this model is that it allows us to also study a situation where the long mode experiences the non-attractor phase (by extending the USR phase, i.e., taking the limit $\tau_i \to -\infty$). In this case, the standard consistency relation Eq.~\eqref{eq:fNL_SR} is violated but, we claim, the more general  one Eq.~\eqref{eq:fNL_final} holds. 
 
 This model has  already been discussed in \cite{Cai:2022erk} but the bispectrum is only obtained using the $\delta N$-formalism. We will obtain $\fNL(k_s)$ that is valid for all values of $k_s$ including the peaks where we show that it  generically differs from the results of the $\delta N$-formalism. The model without the step is also extensively discussed in the literature, see e.g., \cite{Chen:2013aj}. The full expression for $\fNL(k_s)$ is studied in this simplified situation in \cite{tada2023cancellation} but not for the case of extended USR. Below, we only outline the existing analysis and refer the readers to e.g.,  \cite{Cai:2022erk} and  \cite{tada2023cancellation} for further details but also provide some details in Appendix~\ref{app:in-in}. While we keep the analytic calculations generic throughout this section, when we present numerical results, we choose $g=1$ which corresponds to the case without the step. This is because, as mentioned earlier, an infinitely sharp step is unrealistic and one expects the slope of the step to be observable to sufficiently short modes. The choice of other parameters will be reported in the figures.  
 
\subsection{Transient ultra-slow-roll inflation}
\label{sec:example_transient}

\begin{figure}
    \centering
    \includegraphics[scale=0.5]{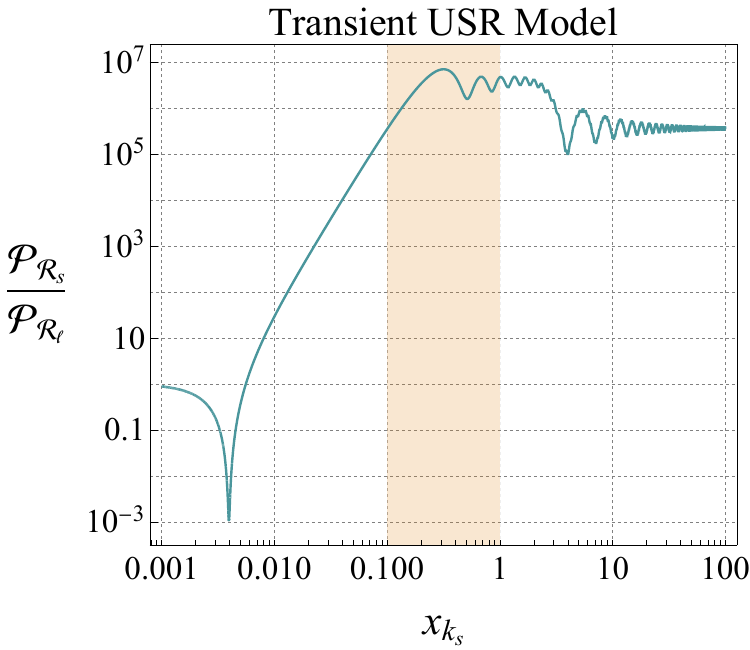}
    \hspace{.4cm}
    \includegraphics[scale = 0.5]{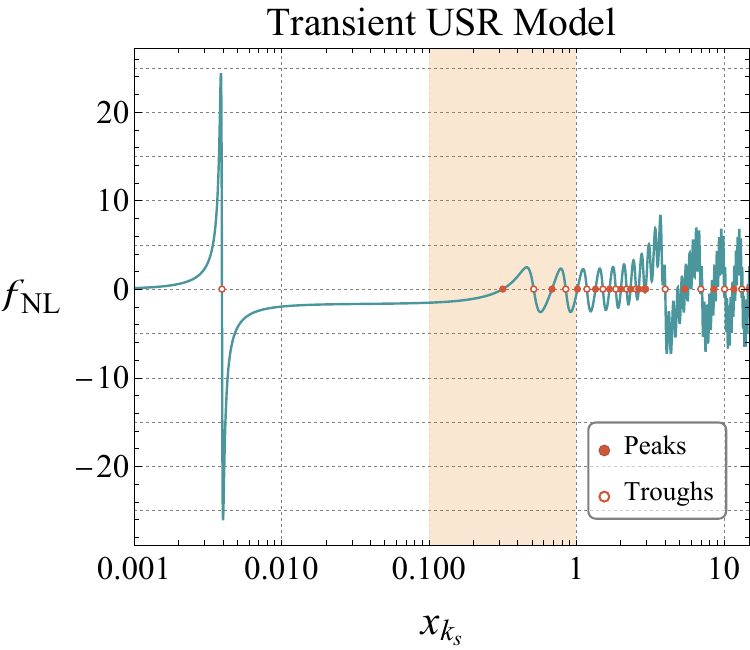}
    \caption{The power spectrum (left) and the squeezed limit bispectrum (right) for the transient USR model as functions of $x_{k_s}=-k_s \tau_e$. The shaded regions indicate the modes that exit the horizon during the USR phase. On the right panel, we also mark the peak and trough scales of the power spectrum. For this figure, we set $g=1$,  $h=-10$  and  $\tau_i/\tau_e=-10$. }
    \label{fig:double transitions}
\end{figure}

We start with analyzing the situation where the USR phase is transient, so we have two phase transitions to take into account.
The mode function of the curvature perturbation can be expressed generally  by
\begin{equation}
\label{eq:mode_func}
\mathcal{R}_k(\tau)= \frac{H}{\sqrt{4 \epsilon(\tau) k^3}} \left[ \alpha_k(1+i k \tau) e^{-i k \tau}+\beta_k  (1-i k \tau) e^{i k \tau} \right],
\end{equation}
where $H$ is the Hubble parameter and is assumed to be a constant while the time dependence of $\epsilon$ is arbitrary. $\alpha_k$ and $\beta_k$ are the Bogolyubov coefficients that may be determined at each phase by matching the mode function to the one from the previous phase at the transition time, using the continuity of $\calR$ and $\epsilon \calR'$ \cite{Namjoo_2012}. The coefficients at the first SR phase are also fixed by the Bunch-Davies vacuum. We leave the details of the derivation of the Bogolyubov coefficients to the  Appendix.~\ref{app:Mode Functions} and only present here the late time power spectrum which is given by
\begin{equation}
    \label{eq:power spectrum1}
    \mathcal{P}_{\mathcal{R}}(k) = \dfrac{ H^2}{8 \pi^2 \epsilon_V} |\alpha_k^{(3)} + \beta_k^{(3)}|^2 \ ,
\end{equation}
where $\alpha^{(3)}_k$ and $\beta^{(3)}_k$ are the Bogoliubov coefficients at the third stage of the evolution (i.e., the final SR phase) which are presented in Eq.~\eqref{eq:alpha3 and beta3} of  Appendix.~\ref{app:Mode Functions}. Note that the dependence of $  \mathcal{P}_{\mathcal{R}}$ on the history of the evolution is via the explicit appearance of $\tau_i$ and $\tau_e$ in the Bogoliubov coefficients as well as the dependence of $h$ and $g$ on $\epsilon_e = \epsilon_i (\tau_i/\tau_e)^6$. It is clear from Eqs.~\eqref{eq:alpha2 and beta2} and \eqref{eq:alpha3 and beta3} and from the functional form of  $\epsilon_e$ that the power spectrum can be entirely written in terms of relative scales, as claimed in Sec.~\ref{sec:general_cosistency}. The left panel of Fig.~\ref{fig:double transitions} shows the behavior of the power spectrum.

In the transient USR scenario,  the long mode leaves the horizon before the first phase transition and is completely frozen during the non-attractor evolution, hence the standard consistency relation Eq.~\eqref{eq:fNL_SR} holds. The validity Eq.~\eqref{eq:fNL_SR} is already confirmed in Ref.~\cite{tada2023cancellation} in a similar but simpler setup (no step is considered). So we do not show the explicit comparison that we have made (including the step) between the consistency relation and the in-in calculations of this model (but we will do so for the more complex situation of Sec.~\ref{sec:example_extended}).
The right panel of Fig.~\ref{fig:double transitions} shows  the behavior of $\fNL$ according to Eq.~\eqref{eq:fNL_SR}, where we have also marked the extrema of the power spectrum. Clearly, $\fNL$ vanishes at the peaks and troughs. 

\subsection{Extended ultra-slow-roll inflation}
\label{sec:example_extended}

\begin{figure}
    \centering
    \includegraphics[scale=.5]{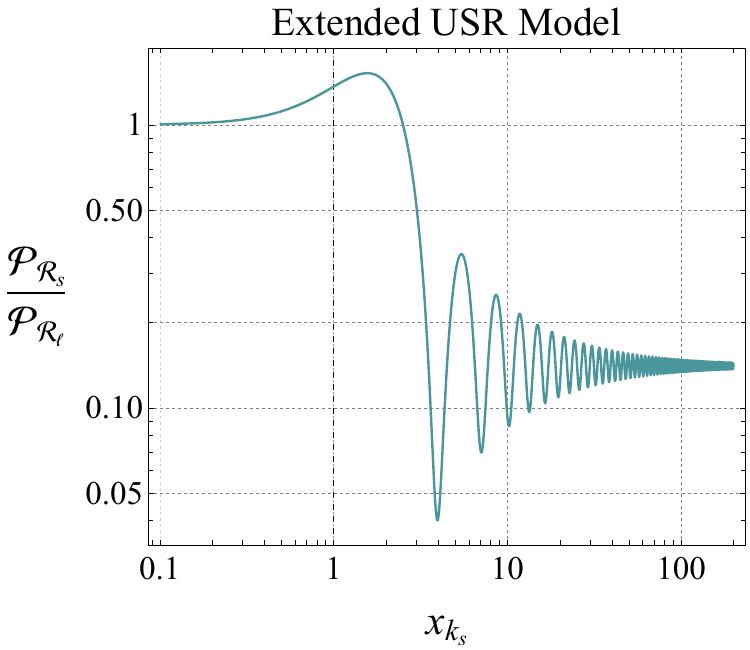}
    \hspace{.5cm}
    \includegraphics[scale=.5]{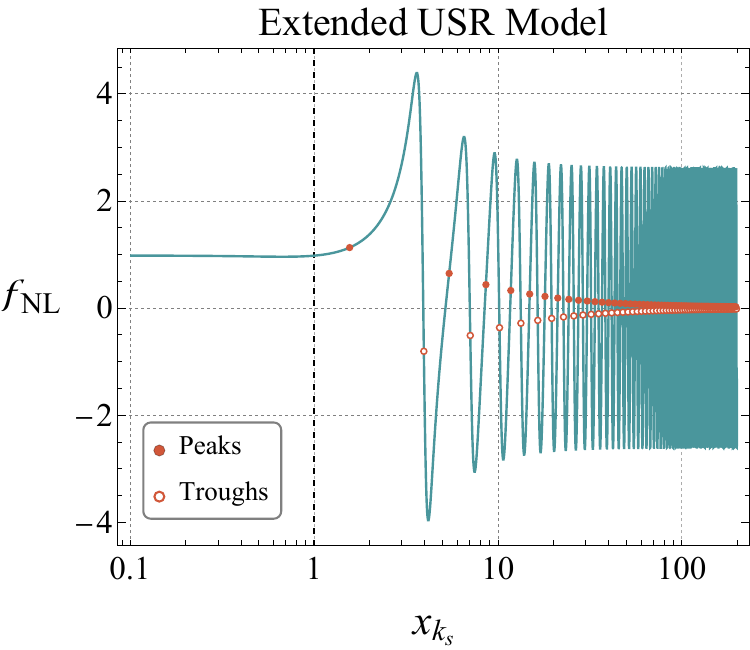}
    \caption{The power spectrum (left) and the squeezed limit bispectrum (right) for the extended USR model ($\tau_i \to -\infty$). On the right panel, we also mark the peak and trough scales of the power spectrum. For this figure we used $g=1$ and $h=-10$ (similar to Fig.~\ref{fig:double transitions}).}
    \label{fig:single transition}
\end{figure}

A more non-trivial exercise for the application of our consistency relation is  the case where the USR phase is extended so that the long mode leaves the horizon in the USR phase. In this case, we have only one phase transition (from USR to SR) and the power spectrum can be obtained by taking the limit $\tau_i \to -\infty$ of Eq.~\eqref{eq:power spectrum1} which results in
\begin{equation}
    \begin{aligned}
        \label{eq:power of the single transition}
        \mathcal{P}_{\mathcal{R}} = \frac{ H^2}{32 \pi^2 g^2  x_k^6 \epsilon_V} |\Gamma|^2\ ,
    \end{aligned}
\end{equation}
where $x_k=-k \tau_e$ and 
\begin{eqnarray}
\label{eq:Gamma}
\Gamma &\equiv& \left(6 - g^2 (6 + h) (1 - i x_k) - 2 x_k (3 i + x_k)\right) \left(\sin(x_k)-x_k \cos(x_k)\right) \\ \nonumber 
&&+ 2 g^2  x_k^2 (1 -i x_k)  \sin(x_k).
\end{eqnarray}
The left panel of Fig.~\ref{fig:single transition} visualizes this power spectrum. 

Let us now study the squeezed limit bispectrum by applying our consistency relation Eqs.~\eqref{eq:fNL_interm} or \eqref{eq:fNL_final}. We need to inspect how the power spectrum $\calP_{\calR_s}$ (Eq.~\eqref{eq:power of the single transition} with the replacement $k\to k_s$) depends on the history of its evolution. It is clear from Eq.~\eqref{eq:power of the single transition} that the final power spectrum is sensitive to $\tau_e$ so the effect of the long mode on the transition time must be taken into account. Clearly, there is no dependence on the horizon-crossing time. Moreover, one can argue that the  contribution from the initial conditions can also be neglected: The power spectrum Eq.~\eqref{eq:power of the single transition} is affected by the initial conditions via the slow-roll parameter $\epsilon =( \dot \phi_{0}^2/2H^2) (\frac{\tau}{\tau_{0}})^{6}$ where $\dot \phi_0$ is the initial inflaton's velocity and $\tau_{0}$ is some initial conformal time  (i.e., we assume $|\tau_\ell| < |\tau_{0}| \ll |\tau_e|$).\footnote{As discussed in Sec.~\ref{sec:general_cosistency} the choice of $\tau_{0}$ is rather arbitrary and we only require it to be early enough so that an even earlier effect of the long mode can be neglected. Let us stress that this expression for $\epsilon$ is different from the one for the transient USR scenario where we had $\epsilon=\epsilon_i (\tau/\tau_i)^6$ in the USR phase where $\tau_i$ is the conformal time at the onset of the USR phase and $\epsilon_i $ is the attractor slow-roll parameter prior to the USR phase.} In principle, $\calR_\ell$ can affect both $\dot \phi_{0}$ and $\tau_{0}$. The former is negligible since the induced field $\Delta \phi = -(\dot \phi/H) \calR_\ell$ is almost conserved even in the USR regime so that it cannot contribute significantly to the inflaton's velocity. The effect of $\calR_\ell$ on $\tau_{0}$ can also be neglected since $r(\tau)$ at $\tau_0$ is much smaller than that at later times due to the rapid growth of the power spectrum.

We are thus left with the only contribution  from the impact of $\calR_\ell$ on $\tau_e$. We need to know $r(\tau_e)$ (which is $\calR_\ell(\tau)$ at $\tau_e$ relative to that at $\tau\to 0$ ) which can be deduced from the expansion of the mode function Eq.~\eqref{eq:mode_func} for small $k$. This yields $r(\tau_e)=\frac{g^2h}{g^2h-6}$. Then Eqs.~\eqref{eq:fNL_interm} and \eqref{eq:fNL_final} reduce, respectively, to
\ba 
\label{eq:fNL_example}
\dfrac{12}{5}\fNL(k_s) = \big( \frac{g^2h}{6-g^2h} \big)\dfrac{d \ln \calP_{\calR_s}}{d\ln \tau_e} = 
 \big( \frac{g^2h}{g^2h-6} \big) \left[ \dfrac{\partial \ln \calP_{\calR_s}}{\partial \ln x_0}- (n_s-1)  \right]
\ ,
\ea 
where, for the last equality, we noted that $\tau_e$ appears in the formulation of Eq.~\eqref{eq:fNL_final} via $x_{k_s}=-k_s \tau_e$ and $x_0=\tau_0/\tau_e$ and we neglected $r(\tau_0)$ relative to $r(\tau_e)$ in the last term of Eq.~\eqref{eq:fNL_final}. Note that  $g$ and $h$ in Eq.~\eqref{eq:power of the single transition} also depend on $\tau_e$ through $\epsilon_e \propto \tau_e^{6}$ (and recall that $\epsilon_s=\epsilon_e-3\kappa$).
Remarkably,  this result coincides with the one from explicit in-in calculations. The in-in calculations are rather tedious but we outline the main steps in  Appendix.~\ref{app:in-in} where we also present the explicit form of $\fNL$ (Eq.~\eqref{eq:fnl in-in}).
The right panel of Fig.~\ref{fig:single transition} shows the behavior of $\fNL$ as a function of $k_s$ where we also mark the location of peaks and troughs.  

Let us study the analytic form of $\fNL$ in the two  limits of interest.  We first take the limit $k_s \to 0$ which corresponds to the modes that are super-horizon when the transition to the SR phase occurs. We have 
\ba 
\label{eq:fNL_deltaN}
\dfrac{12}{5} \fNL(k_s) \to  \dfrac{6 h(g^4 h+6 g^2-6)}{(g^2h-6)^2}
\qquad \text{for $k_s \to 0$}\, .
\ea 
Note that, as stated in Sec.~\ref{sec:limits}, this is the limit where  the $\delta N$-formalism is also valid and we reproduced precisely that result \cite{Cai:2022erk}.  
The same result is also obtained in Ref.~\cite{firouzjahi2023sign} but it is claimed that it holds for the peak. Our expression for the peak would be different since $k_s \to 0$ is not a valid limit for the first peak mode which indeed leaves the horizon after the phase transition (i.e., $x_{k_s} >1$ for the first peak; see the left panel of Fig.~\ref{fig:single transition}). Crucially, contrary to the claim of Ref.~\cite{firouzjahi2023sign}, it is possible to have a negative $\fNL$ according to Eq.~\eqref{eq:fNL_deltaN} (for an appropriate range of $g$ when $|h|<6$) whereas our full result Eq.~\eqref{eq:fNL_example} implies that at the peaks $\fNL$ always remains  positive (see the right panel of Fig.~\ref{fig:single transition}).\footnote{We note that Ref.~\cite{firouzjahi2023sign} assumes continuity of $\dot \calR$ at the phase transitions for derivation of a consistency relation which does not hold in this setup. More generally, and for the scenario we consider here (which is the same as the one considered in Ref.~\cite{firouzjahi2023sign}), $\epsilon\dot \calR$ is the continuous quantity at the transitions.} This is because all the assumptions discussed in Sec.~\ref{sec:sign} for the positivity of $\fNL$ at the peaks are satisfied in this example.

In the opposite  limit (i.e.,  $k_s \to \infty$) , Eq.~\eqref{eq:fNL_example} (or, more explicitly, Eq.~\eqref{eq:fnl in-in}) reduces to
\ba 
\dfrac{12}{5} \fNL(k_s)\to   \big( \frac{g^2h}{6-g^2h} \big) (n_s-1) \ , \qquad \text{for $k_s \to \infty$}\ ,
\ea 
where in the same limit we have 
\ba 
 (n_s-1) \to 
\begin{cases}
	\frac{2 g^2 \left(g^4-1\right) h  x_{k_s} \sin (2  x_{k_s})}{\left(g^2 h-6\right) \big(\left(g^4-1\right) \cos (2  x_{k_s})-g^4-1\big)}  \, , \quad \text{for $g \neq  1$}\, ,
	\\
	\\
	\frac{h^2 \cos (2 x_{k_s})}{h-6} \, , \hspace{3.3cm} \text{for $g=1$}\, .
\end{cases}
\ea 
Despite the significant difference in the behavior of $(n_s-1)$ for $g=1$ (no step) and $g\neq 1$,  the relation $\fNL \propto (n_s-1)$ remains universal, consistent with our generic argument in Sec.~\ref{sec:limits}. 

From the short wavelength expansion of Eq.~\eqref{eq:fNL_example} and keeping next-to-leading order terms, one can also deduce that $\fNL(k_{\text{peaks}}) \simeq  - \fNL(k_{\text{troughs}}) \to 5h|g^2-1|/2(g^2h-6)$  which is a constant (and zero for $g=1$) and the same for all peaks and troughs in this limit. We see once again that $\fNL$ is non-negative at the peaks and non-positive at the troughs.  Finally, note that it is clear from Eq.~\eqref{eq:fNL_example} that $\dfNL>0$ at all peaks  and  $\dfNL<0$ at all troughs as claimed in Sec.~\ref{sec:slope}. 
\\

We have accomplished our main objective in this paper which is to derive and confirm a generalized consistency relation, applicable to the single-field but not necessarily scale-invariant models of inflation and studying its consequences for the peaks. In the next two subsections, we make some side remarks about the implications of the consistency relation on the predictions of the PBH and GW production and leave a thorough analysis for future works. 

\subsection{Probability of large fluctuations}
\label{sec:beta}

We have seen that  $\calP_{\calR_s}$ and $\fNL$ peak at different scales. This raises the question of which scales exhibit the highest probability of the realization of large fluctuations which is of  great importance for the estimation of the mass and abundance of PBHs. The full answer to that question requires non-perturbative knowledge of the probability distribution of smoothed density contrast in real space \cite{Hooshangi:2021ubn,Ezquiaga:2019ftu,Biagetti:2021eep,Musco:2018rwt}. In contrast,   our approach  only offers partial and perturbative access to the correlation functions in Fourier space. In principle, it is possible to obtain real space (but still perturbative) results if the full bispectrum is calculated but it will be a complex task due to the high scale dependence and is beyond the scope of this paper.  In this section, we only try to make a rough estimate of the effect of non-linearities by studying  approximations to the full probability distribution. Our results should be considered as speculations,  motivating a more careful analysis. 

We start with defining a quantity $\beta(k_s)$ as follows
\ba 
\label{eq:beta}
\beta(k_s) \equiv \int_1^{\infty} \rho_{\hat \calR} (k_s) \, d\hat \calR \ ,
\ea 
where $\hat \calR$ is a random variable (to be defined) with the probability distribution function $ \rho_{\hat \calR}$. $k_s$ may be understood as the variable that determines the smoothing comoving scale of fluctuations in real space   $R_{\text{smooth}} \simeq 1/k_s$. According to that definition, $\beta(k_s)$ is the probability of realization of  the random variable $\hat \calR$ larger than a certain threshold (which is chosen to be 1) at scale $R_{\text{smooth}} $.  

Now we need to determine a probability distribution for $\hat \calR$. A natural relation for the random variable $\hat{\calR}$ that takes into account the leading order non-linearities may be expressed by
\ba 
\label{eq:Rhat_NG}
\hat \calR \equiv  \calR_g+\frac{3}{5} f^{\text{eff}}_{\text{NL}}(k_s) \left( \calR^2_g - \sigma^2(k_s) \right).
\ea 
where $\calR_g$ is a Gaussian random variable with mean zero and variance $\sigma^2(k_s)=\calP_{\calR_s}(k_s)$ and $f^{\text{eff}}_{\text{NL}}(k_s)$ is an effective size  of non-linearity which shall be discussed shortly. This relation  resembles the  expression for the local non-Gaussianity in real space \cite{Komatsu:2001rj} which studying its effect on the probability distribution is straightforward \cite{Byrnes:2012yx}. An expression like this (and with $f^{\text{eff}}_{\text{NL}} = \fNL$) is justified in a scale-invariant scenario but it is also used in models that predict strong violation of scale invariance \cite{Cai:2018dig}. Since we have only access to the squeezed limit bispectrum by the consistency relations and, in the scenarios of interest, its scale dependence is very different from the scale dependence of the power spectrum, it may be hard to justify Eq.~\eqref{eq:Rhat_NG}. However, we try to roughly compensate these defects by choosing an appropriate expression for $f^{\text{eff}}_{\text{NL}}$. To achieve that, we require Eq.~\eqref{eq:Rhat_NG} to roughly give the same skewness to the probability distribution of the field as it is generated by a non-zero three-point correlation function.  We assume a sharp, top-hat window function in Fourier space which significantly suppresses the contributions from the modes with momenta greater than $k_s$ \cite{LoVerde:2007ri,Franciolini:2018vbk}. In this case, the main contribution to the skewness comes from the squeezed configuration but due to  the large difference between the long and the short power spectra, $f^{\text{eff}}_{\text{NL}}$  receives an extra factor as follows
\ba 
\label{eq:fNL_eff}
f^{\text{eff}}_{\text{NL}}(k_s) \simeq   \big(\dfrac{\calP_{\calR_\ell}}{\calP_{\calR_s}} \big) \fNL(k_s). 
\ea 
In this relation,  possible ${\cal O}(1)$ factors are omitted but the prefactor is fixed by requiring the equality $f^{\text{eff}}_{\text{NL}}=\fNL$ in the scale-invariant limit. A more accurate result consists of integrals over long and short modes which are neglected here. This approximation can be roughly  interpreted as considering the effect of a narrow range of  long modes on a narrow range of  short modes responsible for the PBH formation. The factor behind $\fNL$ is reasonable because a long mode with a smaller amplitude is expected to have a smaller effect on the short modes for a fixed strength of correlation.

According to Eq.~\eqref{eq:fNL_eff}, $f^{\text{eff}}_{\text{NL}}$ is suppressed significantly compared to $\fNL$ in the transient USR model due to the large enhancement of the short modes. This suppression makes the non-linear term in Eq.~\eqref{eq:Rhat_NG} ineffective; the resulting $\beta$ would be very close to the Gaussian case.   On the other hand, in the extended model, the peak mode and the long mode are both enhanced by the USR phase  so that $f^{\text{eff}}_{\text{NL}}\sim \fNL$  around the peak. Due to that distinction, in this section, we only study the extended model which is more interesting concerning non-linearities. (However, we also note that this is only the partial result and it would be interesting to study the full effect of non-linearities in both models.) Crucially,  for the extended model, $f^{\text{eff}}_{\text{NL}}$ becomes larger than $\fNL$ at scales shorter than the first peak since $\calP_{\calR_s}<\calP_{\calR_\ell}$ at these scales (see Fig.~\ref{fig:single transition}). 

\begin{figure}
    \centering
    \includegraphics[scale=.5]{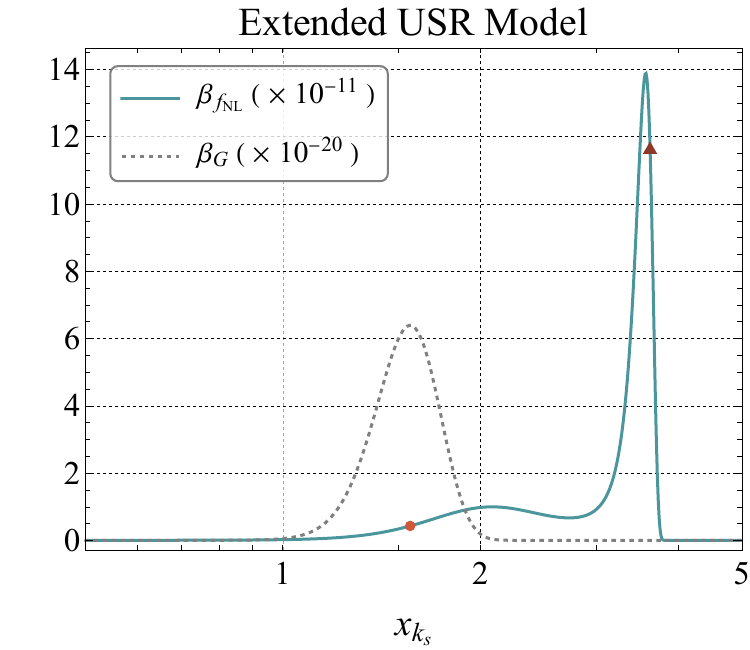}
    \caption{The probability of larger-than-one fluctuations $\beta$ as a function of scale for the  extended  USR model. Note that we multiplied $\beta_G$ and $\beta_{\fNL}$   by different factors to  visualize them simultaneously. We used the same power spectrum   as depicted in  \ref{fig:single transition} (with $g=1$ and $h=-10$) and used  $\calP_{\calR_\ell}= 8 \times 10^{-3}$. The small circle and  triangle on the blue curve indicate the location of the first peak of $\calP_{\calR_s}$ and $\fNL$, respectively. (The peaks of $\fNL$ and $f^{\text{eff}}_{NL}$ occur at $x_{k_s}\simeq 3.62$ and $x_{k_s} \simeq 3.77$, respectively.)}
    \label{fig:beta}
\end{figure}

Fig.~\ref{fig:beta} depicts the behavior of $\beta_{\fNL}$ --- which is $\beta$ as defined in Eq.~\eqref{eq:beta} using the relation \eqref{eq:Rhat_NG} ---  as a function of $k_s$ for the extended USR model. For comparison, we also show the behavior of $\beta_G$ which is the result of the linear theory ($\fNL \to 0$).  We observe that  the  presence of non-linearities has three crucial effects: (i) it increases the probability of large fluctuations, (ii) it shifts and broadens  the window of the PBH formation associated with the first peak and  (iii) a new window opens up due to the large $f^{\text{eff}}_{\text{NL}}$ at shorter scales. While (i) is already a well-known effect, to our knowledge,  (ii) and (iii)  seem to be overlooked in the previous literature. We may expect (ii) and (iii) to lead  to  a non-monochromatic mass spectrum of  PBHs and also cause the mass of the most dominant PBHs  to be different from what is predicted  by the linear theory. In Sec.~\ref{sec:slope} we have argued that  $\dfNL>0$ at the peak. Thus,  the size of $\fNL$ (which is positive according to the discussion of Sec.~\ref{sec:size}) is larger after the peak than it is before the peak. This implies that the presence of non-Gaussianity must shift the window of large $\beta$ from around the peak of the power spectrum to the shorter scales. This generic prediction of the consistency relation is clearly confirmed in this example as is visualized in Fig.~\ref{fig:beta}.\footnote{As another interesting consequence of the consistency relations, consider the situation  where the slow-roll consistency relation Eq.~\eqref{eq:fNL_SR} holds in which case $\fNL$  vanishes at the peak. Since $\dfNL>0$ at the peak, $\fNL$ is negative slightly before and positive slightly after the peak. This may lead to a suppression (enhancement) of $\beta$ at scales slightly larger (shorter) than the peak. This is a situation where the presence of non-linearities can have a dual effect; it causes the window of large $\beta$ to shrink from the left and expand to the right.}

As a further remark related to (iii), let us point out that  it is also possible that non-linearities open up new windows for  PBH production which might be  missed in the perturbative analysis. As an example, consider large scales ($k_s \to 0$) where  scale invariance is restored  and  the validity of the $\delta N$-formalism is justified. Using the $\delta N$-formalism in its non-perturbative form, one can compute $\beta$ non-perturbatively \cite{Hooshangi:2021ubn}. (See Ref.~\cite{Cai:2022erk} for the explicit $\delta N$ relations for the model under consideration.) For the same choice of parameters as for Fig.~\ref{fig:beta}, one obtains $\beta \simeq 2.5 \times 10^{-8}$ from this non-perturbative result, whereas at at the same scales we have $\beta_{\fNL} \simeq 10^{-15}$ and $\beta_G\simeq 2.5 \times 10^{-29}$. This shows the significance of non-perturbative effects.  Furthermore, comparing this non-perturbative result  with the numbers in Fig.~\ref{fig:beta} suggests  that an abundant production of PBHs  is possible at large scales, far away from the peak scale of the power spectrum.   

We conclude that estimating the mass and abundance of PBHs from studying the first peak of the power spectrum may not always be justified. 
However, as stated before, we note that our analysis is oversimplified in various ways.  A thorough study of the PBH formation in such scenarios remains an interesting, open problem.

\subsection{Connected and disconnected trispectra}
\label{sec:GW}

It would also be interesting to study the effect of largely scale-dependent power spectrum and bispectrum on the generation of stochastic GWs. We continue to be speculative, make an order of magnitude estimate of the effect and leave a complete analysis to future work. 

\begin{figure}
	\centering
	\includegraphics[scale=.55]{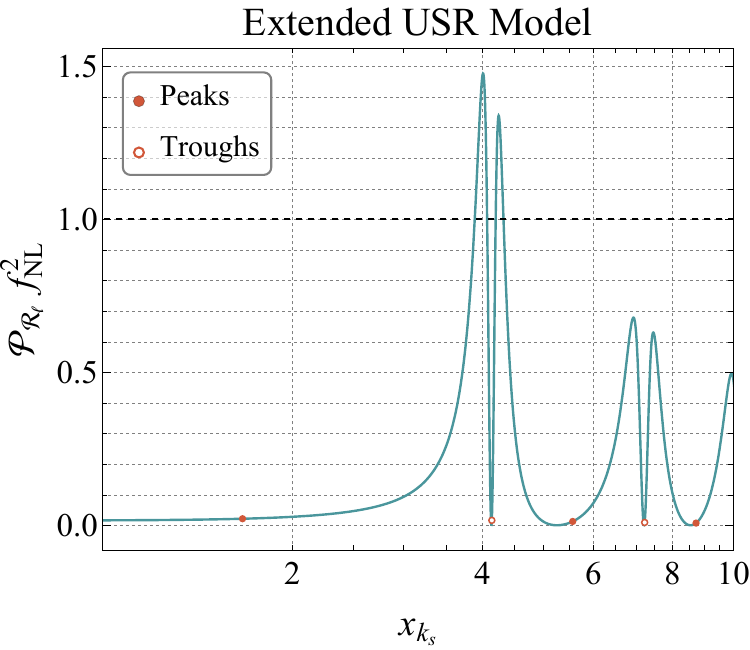}
	\caption{The behavior of $ \calP_{\calR_\ell} f^2_{NL}$ (the contribution of non-linearities to the induced GWs, relative to the linear contribution) as a function of scale for the  extended  USR model. For this figure, we used $g=1$,  $h=-10$ and   $\calP_{\calR_\ell}= 8 \times 10^{-3}$.} 
	\label{fig:GW}
\end{figure}

The amplitude of the induced tensor mode is sourced by the second order scalar perturbation. Therefore, the corresponding energy density of the induced GWs is controlled by the four-point correlation function --- i.e., the trispectrum --- of $\calR_s$. There are two contributions to the four-point function both of which may contribute to the induced GWs, namely, the disconnected and connected trispectrum. The disconnected trispectrum originates from the linear theory whereas the connected one requires non-linear interactions. If the two contributions become comparable, the effect of non-linearities becomes non-negligible \cite{Cai:2018dig}. As an order of magnitude comparison, we consider  $\calP_{\calR_s}^2$ as the size of the disconnected contribution and  $\calP_{\calR_s}^2 \calP_{\calR_\ell} f^2_{NL}$ as the size of the connected contribution. The latter estimation may be justified by considering a connected diagram with four  $\calR_s$ as the external legs and $\calR_\ell$ as the mediator with a cubic long-short interaction with strength $\fNL$. Comparing the two pieces, we roughly conclude that if $\calP_{\calR_\ell} f^2_{NL}(k_s) \gtrsim 1$, the effect of non-linearities cannot be neglected at the comoving scales around $1/k_s$. Fig.~\ref{fig:GW} illustrates the behavior of this factor for a specific choice of  parameters. Notably, this figure suggests that, if the squeezed limit bispectrum is to be important, it ought to be so at scales  different from the first peak of the power spectrum.  We stress  that the full result beyond this order of magnitude estimation requires the full trispectrum at all configurations (see e.g., \cite{Ragavendra:2021qdu,Zhang:2021vak}), the calculation of which is out of the scope of this paper. 

\section{Conclusion}

In this paper, we adapted the general single-field consistency condition advocated in Ref.~\cite{Namjoo_2023} (Eq.~\eqref{main}) for the purpose of analyzing situations where significant deviation from scale invariance occurs. The main motivation for such scenarios is to generate a large abundance of primordial black holes (PBHs) or observable stochastic gravitational waves (GWs). Our main relations are Eqs.~\eqref{eq:fNL_interm} and \eqref{eq:fNL_final} which have important implications for both phenomena. Our consistency relation allowed us to make general and model-independent statements about the  squeezed limit bispectrum at different scales and, in particular, about the size,  sign and  slope of $\fNL$ at the peaks of the power spectrum.

We have argued  that  non-linearities are capable of shifting or expanding the range of scales responsible for the GW  and PBH production and may also open up new windows for both phenomena. The full answer to the question of which range of scales plays the main role in generating PBHs and GWs  requires a more complete analysis and remains an open problem. Specific statements on this matter are likely to be model-dependent, thus studying explicit examples such as the ones considered here or in Refs.~\cite{Hooshangi:2023kss,Domenech:2023dxx,Pi:2022ysn,Cai:2022erk,Cai:2021zsp,Ragavendra:2020sop,Lin:2020goi,Saito:2008em,Gao:2020tsa} would be fruitful.

Our general consistency relation shows that the value of $\fNL$ at the largest peak of the power spectrum can be different from the predictions of the $\delta N$-formalism or any other methods that rely on the gradient expansion for the peak scale \cite{firouzjahi2023sign,Iacconi:2023ggt}. Thus, claims like that the behavior of the tail as predicted by the $\delta N$-formalism applies to the peak may not be justified \cite{Cai:2022erk}. This may be expected because these methods do not carry scale-dependent information while the situation under consideration is highly scale-dependent. The $\delta N$-formalism may be improved by going beyond the leading order in the gradient expansion \cite{Takamizu:2010xy,Jackson:2023obv,Ozsoy:2019lyy}. This may allow us to study non-linearities at the first peak but not at the next peaks that appear at shorter scales. Crucially, our study of the explicit examples suggests that  the scales shorter than the first peak can  be as important in generating PBHs or GWs due to the enhancement of the tail of the distribution by non-linearities. 

It would be crucial to go beyond the three-point correlation function and it would be interesting to explore the possibility of extending the consistency relation (Eq.~\eqref{main}) for that purpose. In the specific case of four-point function in the collapse limit,  one may expect to have $\tau_{\text{NL}} \propto f^2_{\text{NL}}$ which allows us to make general statements about $\tau_{\text{NL}}$ from what we concluded for $\fNL$ \cite{Ozsoy:2021pws}. We  leave these studies to future works. 

\acknowledgments
We thank Hassan Firouzjahi and Mehrdad Mirbabayi for fruitful discussions. This work is supported by INSF of Iran under the grant number 4022911.

\appendix
\section{Explicit in-in Calculation}
\label{app:in-in}
In this appendix, we provide some details of the model studied in Sec.~\ref{sec:example} and also calculate the  non-Gaussianity of the extended USR model of  Sec.~\ref{sec:example_extended}, using the standard in-in formalism.

\subsection{Background evolution}
\label{app:background}
The background equations of motion in single-field inflation are
\begin{equation}
    \label{eq:Klein-Gordon}
    \Ddot{\phi} + 3 H \dot{\phi} + V_{,\phi}(\phi) = 0 , \quad H^2  = \frac{1}{3 \mpl^2} \big( \frac{1}{2} \dot{\phi}^2 + V(\phi) \big),
\end{equation}
where the overdot  denotes  derivatives with respect  to the cosmic time. Using the number of e-folds ($N \propto \log(a)$) as the clock, the first equation can be rewritten by
\begin{equation}
    \label{eq:Klein-Gordon in N}
    \dfrac{\mathrm{d}^2\phi}{\mathrm{d}N^2} + 3 \dfrac{\mathrm{d}\phi}{\mathrm{d}N} + \dfrac{V^{\prime}(\phi)}{V(\phi)} \simeq 0 \ ,
\end{equation}
where an $\epsilon$-suppressed term  is neglected. 
The slow-roll parameters are defined by
\begin{equation}
    \label{eq:SR definitions}
    \epsilon \equiv -\frac{\dot{H}}{H^2} = \frac{1}{2} \big( \dfrac{\mathrm{d}\phi}{\mathrm{d}N} \big)^2 , \quad \eta \equiv \frac{\dot{\epsilon}}{H \epsilon}. 
\end{equation}

In the transient USR model with the potential as presented in Eq.~\eqref{eq:Potential}, the inflaton field begins its journey from an SR phase and reaches $\phi_i$, the value at the first phase transition,  at the conformal time $\tau_i$. Thus, we have
\begin{equation}
\label{eq:epsilon and eta first SR}
\text{For}\quad \tau \leq \tau_i:\quad \epsilon(\tau) \simeq \epsilon_i,\quad \eta(\tau) \approx 0.
\end{equation}
 Then it transits sharply to the USR phase after which $\eta \simeq  -6$. Thus at the transition, the  SR parameters can be modeled as follows
\begin{equation}
\label{eq:epsilon and eta first transition}
\text{For}\quad \tau_{i-} \leq \tau \leq \tau_{i+}:\quad \epsilon(\tau) = \epsilon_i,\quad \eta(\tau) = -6 \theta(\tau - \tau_i).
\end{equation}
 
 In the USR phase, one can  find the SR parameters by solving Eq.~\eqref{eq:Klein-Gordon in N} for a constant potential,
 \begin{equation}
 \label{eq:epsilon and eta USR}
 \text{For}\quad \tau_i \leq \tau \leq \tau_e:\quad \epsilon(\tau) = \epsilon_i \big( \frac{\tau}{\tau_i} \big)^6,\quad \eta(\tau)  =  -6,
 \end{equation}
 where $\tau_e$ is the comformal time when $\phi_e$ is reached, where the USR phase terminates. Thus the SR parameters just at the end of the USR stage would be $\epsilon(\tau_{e-} )= \epsilon_i \big( \frac{\tau_e}{\tau_i} \big)^6 \equiv \epsilon_e$ and $\eta(\tau_{e-}) = -6$.
 
 After the USR phase, the inflation reaches a sharp step exactly located at $\phi_e$. As a result, the inflaton speeds up or down depending on the sign of $\kappa$ in Eq.~\eqref{eq:Potential}. Denoting  the SR parameter right after the step by $\epsilon_s$, we can parameterize this change of speed with $g \equiv \sqrt{\frac{\epsilon_s}{\epsilon_e}}$.  Around the second phase transition, the first slow-roll parameter is discontinuous resulting in 
 \begin{equation}
 \begin{aligned}
 \label{eq:epsilon and eta second transition}
 \text{For}\quad \tau_{e-} \leq \tau \leq \tau_{e+}:\quad & \epsilon(\tau) = \epsilon_e \theta(\tau_e -\tau) + \epsilon_s \theta(\tau - \tau_e)\ , \\
 & \eta(\tau) = -6 \theta(\tau_e - \tau) - (h+6) \theta(\tau - \tau_e)- \frac{\tau}{\epsilon(\tau)} \delta(\tau - \tau_e)\left( \epsilon_s-\epsilon_e \right),
\end{aligned}
 \end{equation}
 where $h \equiv -6 \sqrt{\frac{\epsilon_V}{\epsilon_s}}$.
 
At the final SR phase, the inflaton's trajectory relaxes to its new attractor. The SR parameters during this relaxation time can be determined by solving  the Klein-Gordon equation Eq.~\eqref{eq:Klein-Gordon in N} with the potential Eq.~\eqref{eq:Potential} for $\phi<\phi_e$:
 \begin{equation}
\label{eq:epsilon and eta relaxation}
\text{For}\quad \tau_e \leq \tau:\quad \epsilon(\tau)=\epsilon_s\left(\frac{h}{6}-(1+\frac{h}{6})\big(\frac{\tau}{\tau_e}\big)^3\right)^2, \quad  \eta(\tau)=-\frac{6(6+h)}{(6+h)-h\left(\frac{\tau_e}{\tau}\right)^3}.
\end{equation} 
 Now with all the SR parameters in hand, we can determine the mode functions for each stage of  inflation.

 \subsection{Mode Functions}
 \label{app:Mode Functions}
The comoving curvature perturbation obeys the following equation
\begin{equation}
    \label{eq:Sasaki-Mukhanov}
    \mathcal{R}_k^{\prime \prime}(\tau)-\dfrac{(2+\eta)}{\tau}\mathcal{R}_k^{\prime}(\tau)+k^2 \mathcal{R}_k(\tau)=0.
\end{equation}
Using Eq.~\eqref{eq:Sasaki-Mukhanov} along with the Bunch-Davies vacuum, one obtains the mode function of the first slow-roll period as follows,
\begin{equation}
    \label{eq:R1}
    \mathcal{R}^{(1)}_k(\tau)=\frac{H}{\sqrt{4 \epsilon k^3}}(1+i k \tau) e^{-i k \tau}.
\end{equation}
To obtain the mode functions for the USR  and the second SR phases, we quantify them using the Bogolyubov coefficients \cite{Cai_2018,Cai:2022erk,Firouzjahi:2023aum}
\begin{equation}
    \label{eq:R2}
     \mathcal{R}_k^{(2),(3)}(\tau)=\alpha^{(2),(3)}_k \frac{H}{\sqrt{4 \epsilon k^3}}(1+i k \tau) e^{-i k \tau}+\beta^{(2),(3)}_k \frac{H}{\sqrt{4 \epsilon k^3}}(1-i k \tau) e^{i k \tau},
\end{equation}
where the superscripts $(2)$ and $(3)$ represent the coefficients for the second (USR) and third (SR) phases of the evolution, respectively. This mode function is an approximate solution of Eq.~\eqref{eq:Sasaki-Mukhanov} as long as $H$ is kept a constant but $\epsilon$ can have an arbitrary time dependence \cite{Cai_2018}. The Bogolyubov coefficients may be determined by the continuity of $\calR$ and $\epsilon \calR'$ at each phase transition \cite{Namjoo_2012} which result in
\begin{equation}
    \label{eq:alpha2 and beta2}
    \alpha_k^{(2)}=1-\frac{3 i}{2 x_k^3 x_i^3}\left(1+x_k^2 x_i^2\right), \quad \beta_k^{(2)}=\frac{3 i}{2 x_k^3 x_i^3}\left(1-i x_k x_i\right)^2 e^{2 i x_k x_i},
\end{equation}
and 
\begin{equation}
\label{eq:alpha3 and beta3}
\begin{pmatrix}
\alpha^{(3)}_k \\
\\
\beta^{(3)}_k
\end{pmatrix}
=
\frac{1}{4 g x_k^3}
\begin{pmatrix}
\gamma_{11} \quad \gamma_{12}\\
\\
\gamma_{21} \quad \gamma_{22}
\end{pmatrix}
\begin{pmatrix}
\alpha^{(2)}_k \\
\\
\beta^{(2)}_k
\end{pmatrix},
\end{equation}
 where $x_k = - k \tau_e$, $x_i = \frac{\tau_i}{\tau_e}$ and
\begin{equation}
\begin{gathered}
\label{eq:gammas}
\gamma_{11} = \gamma^*_{22} = -i \left[ -6 + g^2 (6+h) + \left(-4 + g^2 (4+h) \right) x_k^2 + 2 i (1 + g^2) x_k^3 \right],\\
\\
\gamma_{12} = \gamma^*_{21} = e^{-2 i x} \left(x_k - i \right) \left[ -6 + g^2 (6+h) + i (-6 + g^2 (6+h) ) x_k+ 2 ( 1 - g^2 ) x_k^2 \right]. 
\end{gathered}
\end{equation}
For $g=1$ i.e., when there is no step, $\alpha^{(3)}_k$ and $\beta^{(3)}_k$ reduce to the Bogolyubov coefficients reported in \cite{firouzjahi2023sign}. The final power spectrum is given by
\begin{equation}
    \label{eq: power spectrum}
    \mathcal{P}_{\mathcal{R}} = \dfrac{ H^2}{8 \pi^2 \epsilon_V} |\alpha_k^{(3)} + \beta_k^{(3)}|^2.
\end{equation}

The power spectrum  for the extended USR phase can be obtained by taking the limit $ k \tau_i \to -\infty$ which results in $\alpha^{(2)}_k \to 1, \beta_k^{(2)} \to 0$ and  
\begin{equation}
\label{eq:extended USR Bogolyubov}
\alpha^{(3)}_k = \frac{\gamma_{11}}{4 g x_k^3}, \quad \beta^{(3)}_k = \frac{\gamma_{21}}{4 g x_k^3}.
\end{equation}
Hence, the final power spectrum in the extended USR model is
\begin{equation}
\label{eq:extended power}
\mathcal{P}_{\mathcal{R}} = \dfrac{ H^2}{32 \pi^2 g^2 x_k^6 \epsilon_V}  | \Gamma|^2,
\end{equation}
where 
\begin{equation}
\label{eq:Gamma in terms of gammas}
\Gamma = \frac{e^{-i x_k}}{2} \left( \gamma_{11} + \gamma_{21} \right).
\end{equation}

\subsection{Non-Gaussianity}
\label{app:non-gaussianity}

We are now prepared to calculate the three-point function for the extended USR model. The leading contribution to the three-point function comes from the following term in the cubic action \cite{Namjoo:2012aa} 
\begin{equation}
\label{eq:Hint}
S_3 \supset \int d \tau d^3 x \frac{a^2 \epsilon}{2} \eta^{\prime} \mathcal{R}^2 \mathcal{R}^{\prime},
\end{equation}
which leads to the following expression for the bispectrum
\begin{equation}
\label{eq:bispectrum_eta}
B_{\mathcal{R}}\left(k_1, k_2, k_3\right)=-\Im \mathcal{R}_{k_1}\left(\tau_0\right) \mathcal{R}_{k_2}\left(\tau_0\right) \mathcal{R}_{k_3}\left(\tau_0\right) \int_{-\infty}^{\tau_0} d \tau a^2 \epsilon \eta^{\prime}\left[\mathcal{R}_{k_1}^*(\tau) \mathcal{R}_{k_2}^*(\tau) \mathcal{R}_{k_3}^{* \prime}(\tau)+2\text {perm.}\right].
\end{equation}
We can break down the integral in Eq.~\eqref{eq:bispectrum_eta} into different pieces for different stages  of inflation.
In the USR phase, $\eta=-6$ is constant, and $\eta^{\prime}=0$. Thus,  $H_\mathrm{int}$ in Eq.~\eqref{eq:Hint} is negligible, and there would be no significant non-Gaussianity from the USR phase. Thus, we are left with the effect of the sharp USR-SR transition and the relaxation period after the transition which we calculate their contribution separately.

To compute the bispectrum from the transition ($B^T_{\mathcal{R}}$) --- i.e., the contribution to Eq.~\eqref{eq:bispectrum_eta} from the range $\tau_{e-} \leq \tau \leq \tau_{e+} $ ---  it is hard to work directly with $H_{\text{int}}$ as given in Eq.~\eqref{eq:Hint} mainly due to the existence of terms proportional to $\delta'$. To circumvent this difficulty,  we perform consecutive integrations by part and use the equation of motion Eq.~\eqref{eq:Sasaki-Mukhanov} to remove higher order time derivatives until no term containing delta-functions or its derivatives remains. This procedure is allowed as long as we  keep track of the boundary terms \cite{Arroja:2011yj}. Then, all integrals become regular and vanish in the limit  $\tau_{e-} \to \tau_{e+}$ and we are only left  with the boundary terms. There are three boundary terms as follow 
\begin{equation}
\int_{\tau_e-}^{\tau_e+} \frac{a^2 \epsilon}{2} \eta^{\prime} \mathcal{R}^2 \mathcal{R}^{\prime} = \frac{1}{2} \left[ a^2 \epsilon \left( \eta \mathcal{R}^2 \mathcal{R}^\prime + \tau \mathcal{R}^2 \nabla^2 \mathcal{R} - 2\tau \mathcal{R} \mathcal{R}^{\prime 2} \right) \right]_{\tau_{e-}}^{\tau_{e+}} \ , 
\end{equation}
which result in
\begin{equation}
\begin{aligned}
\label{eq:boundary terms}
B_{\mathcal{R}}^{T} = a^2\Im \biggl\{\mathcal{R}^*_{k_1} ( 0 ) \mathcal{R}^*_{k_2} ( 0 ) \mathcal{R}^*_{k_3}( 0) \big[ \left( \epsilon \eta  \mathcal{R}_{k_1}(\tau) \mathcal{R}_{k_2}(\tau) \mathcal{R}^{ \prime}_{k_3}(\tau)    \right. 
 - \epsilon \tau  k_1^2  \mathcal{R}_{k_1}(\tau)\mathcal{R}_{k_2}(\tau)\mathcal{R}_{k_3}(\tau)    \\  
\left.    -2  \epsilon \tau  \mathcal{R}_{k_1}(\tau) \mathcal{R}^{ \prime}_{k_2}(\tau) \mathcal{R}^{ \prime}_{k_3}(\tau) \right) +2 \mathrm{perms.}   \big]_{\tau_{e-}}^{\tau_{e+}}  \biggl\}.
\end{aligned}
\end{equation}

Next, we need to perform the integration of  Eq.~\eqref{eq:bispectrum_eta} for the relaxation period  (i.e., for the range $\tau_e <\tau \leq 0$).  Using Eq.~\eqref{eq:epsilon and eta relaxation} we obtain
\begin{equation}
\begin{aligned}
\label{eq:bispectrum relaxation}
B_{\mathcal{R}}^R = \Im \left[\frac{-729 i e^{-i (k_1+k_2+k_3) \tau} H^4 \tau_e^6 (6+h)}{g^4 h^2 k_1^3 k_2^3 k_3^3 \epsilon_e \left((6+h) \tau^3 - h \tau_e^3 \right)^3}   \prod_{i=1}^3  \Lambda_{k_i}(\tau)\right]  _{\tau = 0}^{\tau = \tau_e} \ , 
\end{aligned}
\end{equation}
where 
\ba 
\Lambda_{k_i}(\tau) =(\alpha^{(3)}_{k_i} + \beta^{(3)}_{k_i} ) \left( \beta^{(3)*}_{k_i} (i - k_i \tau) +e^{2 i k_i \tau} \alpha^{(3)*}_{k_i} (i + k_i \tau)\right).
\ea 
The final bispectrum is the sum of the  two contributions $B_{\mathcal{R}} = B^T_{\mathcal{R}} + B^R_{\mathcal{R}}$ which, after taking the squeezed limit and using  Eq.~\eqref{eq:fnl}, results in

\begin{equation}
\label{eq:fnl in-in}
\begin{aligned}
\fNL = \Omega & \biggl\{6\left(-6+g^2(6+h)\right)\left(-3-3 g^2+g^4(6+h)\right)+ 12\left(-1+g^2\right)^2\left(1+g^2\right) x_k^6\\
&+2\left(72+5 g^6(4+h)(6+h)-3 g^2(8+5 h)-7 g^4(24+5 h)\right) x_k^2 \\
&+ 4\left(4-g^2(4+h)\right)\left(3+g^2-g^4(4+h)\right) x_k^4\\
& +\biggl[3\left(-6+g^2(6+h)\right)\left(3+3 g^2-g^4(6+h)\right)+2\left(3-g^2+7 g^4-g^6(9+h)\right) x_k^6 \\
&\quad \, \, +\left(36-3 g^2(-8+h)+g^6(6+h)(16+h)-g^4(156+19 h)\right) x_k^2 \\
& \quad \, \, +\left(12+g^2(4-6 h)-8 g^4(19+3 h)+g^6(136+3 h(14+h))\right) x_k^4\biggl]2 \cos(2x_k)\\
& +\biggl[ \left(g^4 (50 h+204)+12 g^2 (2 h+5)-72  -\left(g^6 (h+6) (9 h+32)\right) \right) x_k^2 \\
&\, \, \, \, \, \, \, \, +\left( g^6 (h (h+30)+140)-2 g^4 (7 h+72)+28 g^2-24 \right) x_k^4
  -4 g^2 \left(g^4-1\right) x_k^6\\
 &\, \, \quad - 12 \left(g^6 (h+6)^2-9 g^4 (h+6)-3 g^2 h+18\right) \biggl] x_k \sin(2 x_k) \biggl\} \, ,
\end{aligned}
\end{equation}
where 
\ba 
\Omega \equiv \frac{5 h}{12\left(-6+g^2 h\right)|\Gamma|^2}.
\ea 
Remarkably, the above $f_{NL}$ matches the one we have derived via the consistency relation Eq.~\eqref{eq:fNL_example}.

\bibliographystyle{JHEP}
\bibliography{references}

\end{document}